\documentclass[10pt,twocolumn,letterpaper]{article}

\usepackage{cvpr}              % To produce the CAMERA-READY version
\usepackage{multirow}
\usepackage[dvipsnames]{xcolor}

\newcommand{\z}{\mathbf{z}}

\newcommand{\ours}{\textsc{CARP3D}}

\definecolor{cvprblue}{rgb}{0.21,0.49,0.74}
\usepackage[pagebackref,breaklinks,colorlinks,citecolor=cvprblue]{hyperref}

\def\paperID{20} % *** Enter the Paper ID here
\def\confName{CVPR}
\def\confYear{2024}

\title{Triage of 3D pathology data via 2.5D multiple-instance learning to guide pathologist assessments}

\author{Gan Gao$^{1,}$\thanks{Equal contribution} , Andrew H. Song$^{2,3,\ast}$, Fiona Wang$^{1}$, David Brenes$^{1}$, Rui Wang$^{1}$, Sarah S.L. Chow$^{1}$, \\Kevin W. Bishop$^{1}$, Lawrence D. True$^{1}$, Faisal Mahmood$^{2, 3}$, Jonathan T.C. Liu$^{1}$\\${^1}$University of Washington, ${^2}$Mass General Brigham, ${^3}$Harvard University\\{\tt\small gangao@uw.edu, asong@bwh.harvard.edu, jonliu@uw.edu}
}
\begin{document}
\maketitle
\begin{abstract}
Accurate patient diagnoses based on human tissue biopsies are hindered by current clinical practice, where pathologists assess only a limited number of thin 2D tissue slices sectioned from 3D volumetric tissue. Recent advances in non-destructive 3D pathology, such as open-top light-sheet microscopy, enable comprehensive imaging of spatially heterogeneous tissue morphologies, offering the feasibility to improve diagnostic determinations. 
% While end-to-end automated diagnostic pipelines for 3D pathology are being developed, 
A potential early route towards clinical adoption for 3D pathology is to rely on pathologists for final diagnosis based on viewing familiar 2D H\&E-like image sections from the 3D datasets. However, manual examination of the massive 3D pathology datasets 
% (equivalent to hundreds of gigapixel 2D sections per biopsy) 
is infeasible. To address this, we present $\ours$, a deep learning triage approach that automatically identifies the highest-risk 2D slices within 3D volumetric biopsy, enabling time-efficient review by pathologists. For a given slice in the biopsy, we estimate its risk by performing attention-based aggregation of 2D patches within each slice, followed by pooling of the neighboring slices to compute a context-aware 2.5D risk score. For prostate cancer risk stratification, 
% (low-grade vs. higher grade), 
$\ours$ achieves an area under the curve (AUC) of 90.4\% for triaging slices, outperforming methods relying on independent analysis of 2D sections (AUC=81.3\%). These results suggest that integrating additional depth context enhances the model's discriminative capabilities. In conclusion, CARP3D has the potential to improve pathologist diagnosis via accurate triage of high-risk slices within large-volume 3D pathology datasets. %Finally, we show that $\ours$-predicted high-risk slices align with a pathologist's evaluation of high-risk slices, demonstrating the potential triaging capability of $\ours$.
% Specifically, we employ an intra-slice attention-based multiple-instance learning mechanism to generate feature representations for 2D image sections and an inter-slice pooling module that integrates neighboring slices to assist target image predictions by automatically emphasizing neighboring images containing clinically important contextual information.  
%Finally, we provide a preliminary demonstration that our approach has the potential to improve diagnostic determinations while reducing pathologist workloads.
\end{abstract}    
\section{Introduction}
\label{sec:intro}

Disease diagnosis and characterization rely upon the accurate histological analysis of biopsies and surgical specimens by pathologists~\cite{3dpath_review,otls_surgical}. In conventional histopathology, only a few thin 2D slices are sectioned from these tissue specimens for microscopic evaluation. Despite being regarded as the gold standard for medical diagnostics for over a century, conventional histology suffers from severe undersampling of tissue specimens, where less than 1\% of a biopsy is examined by a pathologist (4 $\mu m$ thick sections from 1 $mm$ diameter core biopsy). Furthermore, isolated cross-sectional views of complex 3D tissue structures can be ambiguous and misleading~\cite{deeper_section1, deeper_section2}. Recent advances in high-throughput 3D microscopy, along with tissue clearing and fluorescence labeling, now enable non-destructive imaging of large tissue volumes for improved disease characterization, including whole biopsies~\cite{3dpath_tech1, 3dpath_tech2, 3dpath_tech3, 3dpath_tech4, frohn20203d,Forjaz2023Three,hong20203d,3dpath_review, gland3d,protocol, wang20243d}. For example, open-top light-sheet (OTLS) microscopy in conjunction with associated tissue-processing and data-processing methods~\cite{otls2, otls3, otls4, protocol} have shown the ability to generate high-quality 3D pathology datasets at various spatial resolutions with comparable quality to slide-based H\&E histology~\cite{atlas1, atlas2}. 

\begin{figure*}[h]
  \centering
   \includegraphics[width=1.0\linewidth]{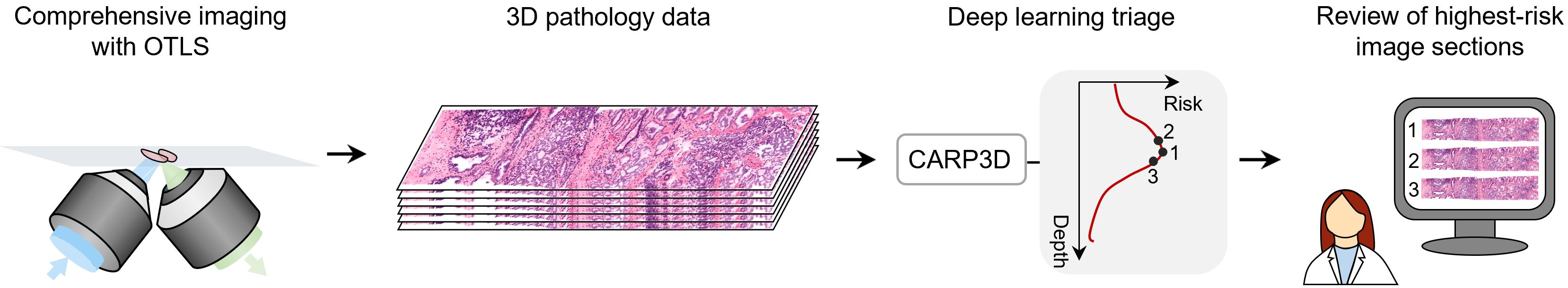}

   \caption{\textbf{Workflow with deep-learning-based triage framework for 3D pathology}. Prostate biopsies are comprehensively imaged in 3D with open-top light-sheet (OTLS) microscopy. A deep-learning-based triage method evaluates all 2D slices within 3D pathology datasets and identifies the highest-risk 2D slices for time-efficient pathologist review.}
   \label{fig:overall}
\end{figure*}

A number of computational methods have been developed to analyze 3D pathology datasets without human intervention~\cite{gland3d,nuc3d,3dweakly}, but fully computational analyses require additional large-scale validation studies before they are ready for clinical deployment. To encourage clinical adoption, it is important to show pathologists 2D cross-sectional images that are false colored to mimic the appearance of histology, allowing pathologists to leverage their training and expertise in interpreting 2D hematoxylin and eosin (H\&E) images~\cite{3dpath_review,atlas2,heter_prostate,prospective24}. However, it is infeasible for pathologists to manually analyze large feature-rich 3D pathology datasets containing hundreds or even thousands of 2D slices. Therefore, there is a need for computational triage methods that can efficiently sift through the vast numbers of slices in each 3D pathology dataset and to identify the highest-risk slices for time-efficient pathologist review (\textbf{Figure ~\ref{fig:overall}}). This approach is a potential low-risk early pathway towards clinical adoption of 3D pathology that keeps pathologists involved in the final diagnosis.
%A number of computational methods have been developed to analyze 3D pathology datasets without human intervention, and have demonstrated the value of 3D vs. 2D pathology~\cite{gland3d,nuc3d,3dweakly}. While fully computational analyses require additional large-scale validation studies before they are ready for clinical deployment, a potential low-risk early path towards clinical adoption of 3D pathology is to keep pathologists involved in the final diagnostic determinations.\cite{3dpath_review,atlas2,heter_prostate,prospective24}. Here, it is important to show pathologists 2D cross-sectional images that are false colored to mimic the appearance of  histology, since this will allow pathologists to leverage their training and expertise in interpreting 2D H\&E images. However, it is infeasible for pathologists to manually analyze large feature-rich 3D pathology datasets containing hundreds or even thousands of 2D image sections.  Therefore, for 3D pathology to be adopted by pathologists, there is a need for computational triaging methods that can efficiently sift through the vast numbers of image sections in each 3D pathology dataset and to identify the highest-risk images for time-efficient pathologist review (Figure ~\ref{fig:overall}). 

For developing a 3D pathology computational framework, it is important to leverage technical advances in 2D computational pathology based on 2D whole-slide images (WSIs), which has witnessed tremendous progress for various clinical outcome prediction tasks such as cancer grading and prognosis, and treatment response predictions \cite{mil1,mil2,mil3}. Due to the gigapixel nature of WSIs, these approaches have centered around a multiple instance learning (MIL) paradigm. In typical MIL setups, WSIs are partitioned into a set of smaller patches (i.e., instances), each of which is encoded into a low-dimensional feature vector using models pretrained on natural images \cite{resnet,imagenet} or more recently, pretrained on in-domain histopathology images \cite{cpath_review,cpath}. These patch-level features are then aggregated with aggregation networks into slide-level representations \cite{maxpaper,attnconf, transmil,cpath_review}. A direct application of MIL frameworks to each 2D slice within a 3D pathology dataset would allow for risk assessment of each slice. However, the analysis of each 2D slice as an independent image does not take advantage of the added context that exists along the 3rd dimension (depth dimension) of a 3D pathology dataset~\cite{3dpath_review}.  For example, a recent work~\cite{3dweakly} has shown that 3D analyses are superior to 2D (in plane) analyses for a patient risk-stratification task based on 3D pathology. However, that approach was designed to make predictions from 3D volumes based on 3D features, making it ill-suited for pinpointing the most important 2D slices at high granularity, which is essential to guide manual review by a human pathologist.

In this work, we propose \textbf{C}ontext-\textbf{A}ware \textbf{R}isk \textbf{P}rediction for 3D pathology (CARP3D), a 2.5D MIL framework for risk prediction. $\ours$ provides a natural mechanism for incorporating contextual information from neighboring slices, which not only leads to enhanced slice-level predictions but also yields a high-resolution (equal to axial sampling pitch) risk profile along the depth axis for triaging applications. Specifically, slices are patched, featurized, and subsequently aggregated into slice-level feature representations by an intra-slice attention-based network in 2D. An inter-slice pooling module subsequently assigns weights to neighboring images based on their diagnostic importance and integrates neighboring features with respective weighting factors to assist predictions on the slice of interest. The attention network allows for the capture of fine-grained details within each slice and the inter-slice pooling module emphasizes contextually relevant information while suppressing irrelevant signals for more accurate assessment of a slice of interest. 
One advantage of our 2.5D approach, as opposed to a full 3D approach, is that it is able to evaluate 3D pathology data at high resolution along the depth axis, and also takes advantage of the emerging family of 2D feature extraction models for 2D pathology images~\cite{cpath, huang2023visual, lu2024visual, chen2024towards}, which are lacking in 3D feature extraction. 
% Similar models for 3D feature extraction are not currently available.

We apply our method to a cohort of prostate cancer biopsies imaged with OTLS, on the task of discriminating between low-grade (Grade group 1) vs. intermediate- to high-grade prostate cancer (Grade group $\geq$ 2), an important clinical task for ensuring that higher-risk patients receive potentially life-saving treatments while low-risk patients are spared serious treatment-related side effects~\cite{GGscore}. 
In terms of predicting risks of given slices, our 2.5D methods outperform the 2D counterparts by a large margin, demonstrating the significance of incorporating 3D contextual information.
% Overall, $\ours$ hints at a clinical potential with enhanced triage of 2D slices within 3D pathology datasets to streamline pathologist review. Our approach shows clinical potential for improved diagnoses over standard-of-care histology, while also saving pathologists’ time. 
The codes for this study are publicly available at \url{https://github.com/alecgao066/CARP3D}.

\section{Related work}
\label{sec:related_work}

\subsection{MIL for 2D WSI classification}

Since directly analyzing gigapixel WSIs often exceeds the capacities of modern GPUs, most classification tasks follow a MIL framework, also termed weakly-supervised learning. In MIL, the WSI is divided into a set of smaller patches (typically $256\times 256$ pixels) from which the low-dimensional features are extracted and aggregated to yield a slide-level feature \cite{cpath_review}. Only a single supervisory label is typically provided for the entire WSI. For patch feature encoding, studies have explored encoding patches by transfer learning with a ResNet50 model pretrained on ImageNet~\cite{maxpaper}, as well as augmentations of patch features in low-data regimes \cite{data_aug}. Recently, CTransPath (a hybrid CNN and Vision Transformer) was trained in a self-supervised learning regime on millions of histopathology images for more generalizable patch features \cite{cpath}. For patch feature aggregation, different approaches have been explored \cite{2dmil1, carmichael2022incorporating}. Among these methods, attention-based networks are gaining popularity, surpassing rule-based models like max, top-k, or average pooling as a result of superior domain adaptability and data efficiency~\cite{maxpaper,2dmil1,2dmil2,2dmil3,2dmil4}. Recent aggregation approaches explore either GNN-based aggregation~\cite{chen2021whole, lee2022derivation} or Transformer self-attention-based aggregations~\cite{transmil,hipt, wagner2023transformer} to more explicitly model relationships between patch features.
% , but at the expense of significantly increased computational demands. In the context of this project, positive labels indicate that the images contain aggressive cancer while negative labels mean the images are entirely free of aggressive cancer. For patch-level analysis, Lu et.al encoded patches into low-dimensional feature vectors by transfer learning with a ResNet50 model pretrained on ImageNet\cite{maxpaper}. Zarrfar et.al further performed data augmentation for patch features to reduce overfitting and to improve model performance in low-data regimes \cite{data_aug}. Wang et.al designed a hybrid CNN and Vision transformer model and pretrained it on millions of unlabeled histopathology data for better and more generalizable patch-feature representations \cite{cpath}

\subsection{Computational 3D pathology to support diagnostic determinations}

In the realm of 3D pathology, where datasets are massive in size, several AI-assisted pipelines have been explored. Some recent studies have focused on 3D segmentation of diagnostically important tissue morphologies, such as prostate glands and nuclei, to facilitate the extraction of hand-crafted features for patient risk stratification \cite{gland3d,nuc3d}. While intuitive and enabling hypothesis testing/generation, the analysis of a small set of intuitive 3D morphological features cannot fully leverage the complex spatial biomarkers contained within 3D pathology datasets, many of which are opaque to human observers. Recent advances in 3D weakly supervised learning allow for “deep features” to be automatically extracted from tissue volumes for patient prognosis~\cite{3dweakly}. Such fully-automated end-to-end deep-learning approaches lack explainability and may initially be challenging for clinicians to adopt. 

Human-in-the-loop strategies that guide pathologists to review important regions could be a low-risk and attractive strategy for more immediate clinical adoption. Towards this end, a fully supervised 2D deep-learning approach was previously developed to identify high-risk image sections within 3D esophageal data for pathologist review \cite{eso_triage}, which was based on a limited number of pixel-level annotations painstakingly provided by a single pathologist. In contrast, $\ours$ is a weakly-supervised learning pipeline trained on diverse image-level annotations that are easily and quickly generated by a panel of pathologists.  Furthermore, it incorporates contextual information in 3D pathology data to achieve better triage performance.

\subsection{2.5D analysis for 3D data}

2.5D analysis considers 3D volumes as a stack of 2D images and leverages neighboring images to refine predictions, incorporating contextual information with lower computational demands than full 3D models. For example, in medical imaging, several models merge nearby gray-scale images into a single multi-channel image to generate segmentation masks or to detect bounding boxes \cite{25d1,25d2,25d3}. Additionally, some approaches adopt auxiliary information extracted from nearby images to improve the 2D model outputs, like adjacent prediction results \cite{infoagg2} or residuals between neighboring images \cite{infoagg1}. Recent works focus on explicitly characterizing context information in 3D datasets with two-stage models, by first conducting 2D analysis for each image and then aggregating features from neighboring images to refine predictions. For such methods, common aggregation approaches include pooling strategies, ensemble methods, or neural networks like RNN \cite{25d1,2step1,2step2,2step3,dbrnn}. However, the majority of these methods were developed for images considerably smaller than 3D pathology images, thus necessitating a 2.5D method tailored for 3D pathology data, such as $\ours$.
\section{Methods}
\label{sec:methods}

\begin{figure*}[h]
  \centering
   \includegraphics[width=1\linewidth]{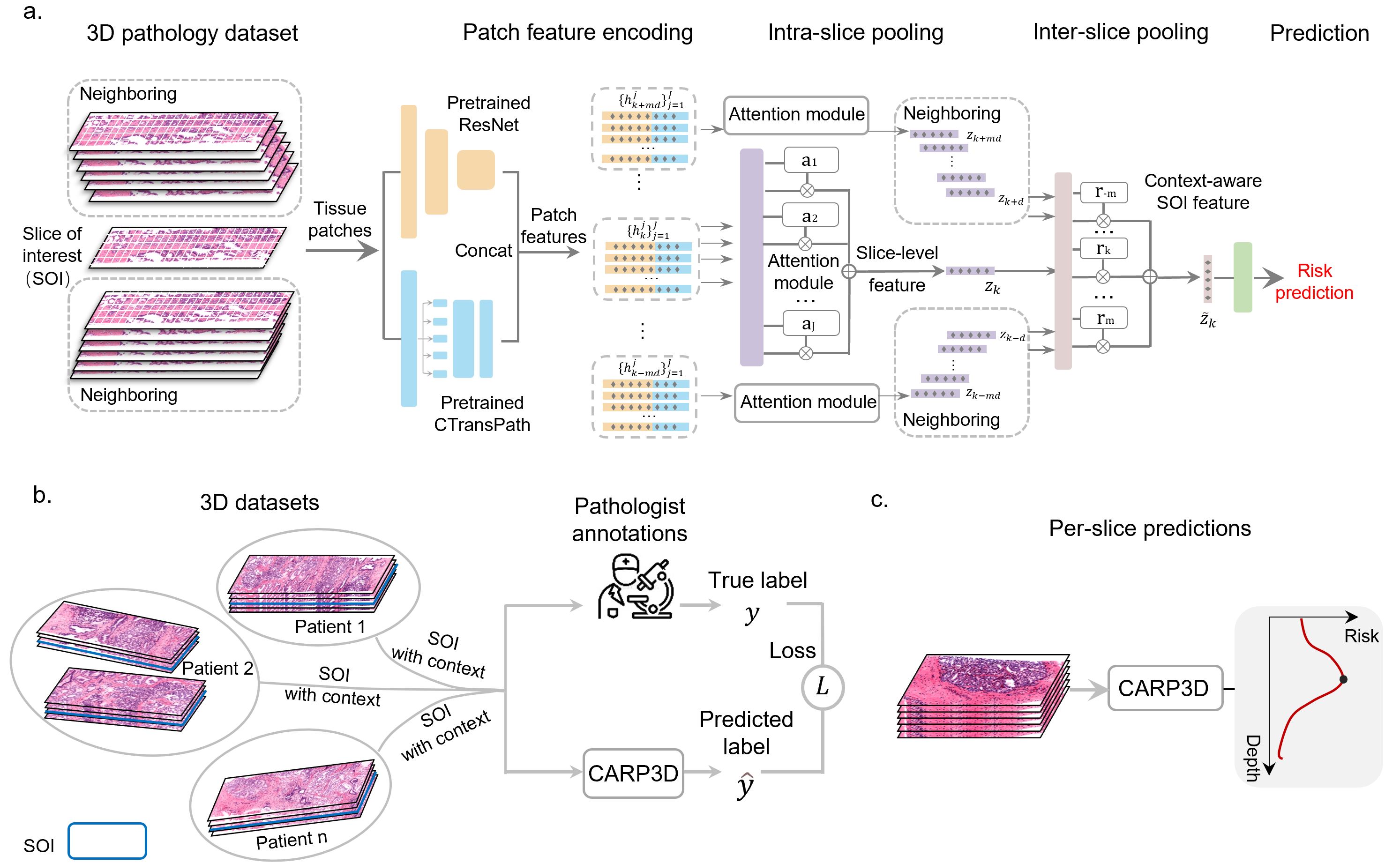}

   \caption{\textbf{$\ours$ architecture}. a) Patches for a 2D slice of interest (SOI) and its neighboring slices are encoded with pretrained ResNet50 and CTransPath. An intra-slice attention module aggregates patch-level features within each slice into slice-level features. Neighboring slice-level features are aggregated through an inter-slice pooling module to produce a context-aware SOI feature for subsequent risk prediction, formulated as a classification task into high- vs. low-risk categories here. b) During training, slices are selected within each 3D sample of the training set, from which the $\ours$ model learns to predict the ground truth labels provided by pathologists. c) Model deployment on 3D pathology data for slice-by-slice risk assessment. The highest-risk slices are selected for pathologist review.}
   \label{fig:model}
\end{figure*}

We represent each 3D pathology image as a stack of 2D images $\{\mathbf{X}_i\}_{i=1}^N$, where $\mathbf{X}_i \in \mathbb{R}^{W \times H \times C}$ with $W$, $H$, and $C$ denoting width, height, and number of channels respectively, and $N$ referring to the number of 2D sections in 3D data. For a slice of interest (SOI) $\mathbf{X}_k$ where $k\in\{1,\ldots,N\}$, we aim to predict the slice-level label $y_k$ (\textbf{Figure~\ref{fig:model}}). For the set of smaller patches from tessellating the SOI and its neighboring slices, we apply a sequence of set aggregation approaches to ultimately construct a single SOI feature for risk prediction. The aim is to harness the extra contextual information provided by the depth dimension: 1) \textit{Intra-slice} attention-based aggregation to construct a slice-level feature and 2) \textit{Inter-slice} aggregation to incorporate neighboring slice-level features to construct a context-aware SOI feature.
% To enhance the accuracy of prediction $\widehat{y}_k$ based on $\mathbf{X}_k$ with additional contextual information, the neighboring sections comprised of images above and below $\mathbf{X}_k$ are incorporated. 
% First, $m$ images above and below $\mathbf{X}_k$ within a certain distance, denoted as $\{\mathbf{X}_{k-m}, ..., \mathbf{X}_k,..., \mathbf{X}_{k+m}\}$, are converted into image-level representations through a sequence of patching, feature extraction, and attention-based multiple-instance learning framework. Then we aggregate adjacent features into image feature at $k$ and generate an image feature with contextual information to predict the class label of $\mathbf{X}_k$. 
 
We first introduce the patch feature extraction (Section~\ref{sec:encoder}) and intra-slice attention-based MIL (ABMIL) (Section~\ref{sec:intra-slice}). We then describe strategies for inter-slice context aggregation from neighboring slice-level features (Section~\ref{sec:inter-slice}), followed by the classification module (Section~\ref{sec:clf}) and training/inference steps (Section~\ref{sec:training}).

%, formulated as a classification task into different risk categories
%-------------------------------------------------------------------------
\subsection{Patch feature encoding}\label{sec:encoder}

We denote the set of patches forming an image $\mathbf{X}_i$ as $\{\mathbf{x}_i^j\}_{j=1}^J$ where $\mathbf{x}_i^j \in \mathbb{R}^{w \times h \times C}$, $w$ and $h$ are lateral dimensions of the patches and $J$ indicates total number of patches in $\mathbf{X}_i$. 
A pretrained feature encoder $f_{\text{enc}}$ then extracts a low-dimensional and representative feature $\mathbf{h}_i^j\in\mathbb{R}^d$ from each patch $\mathbf{x}_i^j$, such that $\mathbf{h}_i^j=f_{\text{enc}}(\mathbf{x}_i^j)$.
% We explored different strategies to encode image patches into low-dimensional feature vectors $\{\mathbf{h}_i^j\}$, where $\mathbf{h}_i^j \in \mathbb{R}^{k \times 1}$ and $k$ corresponds to the encoded feature dimension. We use feature encoder $f: \mathbb{R}^{w \times h \times c} \rightarrow \mathbb{R}^k$ such that $\mathbf{h}_i^j=f(\mathbf{x}_i^j)$, with the feature dimension $k$ determined by the feature encoder choice and whether features from different encoders are concatenated.

To address the domain differences stemming from utilizing pretrained encoders to encode patches of a different data domain (i.e., pretraining dataset: histology, evaluation dataset: OTLS) \cite{maxpaper}, we apply a fully-connected layer with $\operatorname{ReLU}$ nonlinearity to $\{\mathbf{h}_i^j\}_{j=1}^J$ to generate more compressed and domain-specific features $\{\mathbf{\widetilde{h}}_i^j\}_{j=1}^J$, with $\mathbf{\widetilde{h}}_i^j \in \mathbb{R}^{512}$. Specifically, $\mathbf{\widetilde{h}}_i^j=\operatorname{ReLU}(\mathbf{W}^T \mathbf{h}_i^j + \mathbf{b})$, where $\mathbf{W} \in \mathbb{R}^{d \times 512}$ and $\mathbf{b} \in \mathbb{R}^{512}$.

%-------------------------------------------------------------------------
\subsection{Intra-slice attention module}\label{sec:intra-slice}

The fine-tuned patch features $\{\mathbf{\widetilde{h}}_i^j\}_{j=1}^J$ are aggregated by an attention network~\cite{attnconf} into a slice-level feature $\z_i\in\mathbb{R}^{512}$, where the attention score $a_i^j$ computed for each feature $\mathbf{\widetilde{h}}_i^j$ reflects its importance to the final prediction. The attention network is comprised of three sets of parameters $\mathbf{V} \in \mathbb{R}^{512 \times 256}$, $\mathbf{U} \in \mathbb{R}^{512 \times 256}$, and $\mathbf{W} \in \mathbb{R}^{256 \times 1}$ and the attention calculation,
\begin{equation}
  a_i^j = \frac{\exp\{\mathbf{W}^T(\operatorname{tanh}(\mathbf{V}^T \mathbf{\widetilde{h}}_i^j) \odot \operatorname{sigm}(\mathbf{U}^T \mathbf{\widetilde{h}}_i^j))\}}{\sum_{j'=1}^{J} \exp\{\mathbf{W}^T(\operatorname{tanh}(\mathbf{V}^T \mathbf{\widetilde{h}}_i^{j'}) \odot \operatorname{sigm}(\mathbf{U}^T \mathbf{\widetilde{h}}_i^{j'}))\}},
  \label{eq:attnscore}
\end{equation}
where $\sum_{j=1}^Ja_i^j=1$, $\operatorname{tanh}$ and $\operatorname{sigm}$ denote hyperbolic tangent and sigmoid function respectively, and $\odot$ denotes element-wise multiplication operation. The individual patch features are weight-averaged by their corresponding attention scores, resulting in $\z_i$, %$\z_i=\sum_{j=1}^{J} a_i^j \mathbf{\widetilde{h}}_i^j$.
\begin{equation}
  \z_i = \sum_{j=1}^{J} a_i^j \mathbf{\widetilde{h}}_i^j.
  \label{eq:attnsum}
\end{equation}

While $\z_i$ alone is sufficient to predict the slice-level label, we next show that leveraging the contextual information from neighboring slices can improve the performance further, which is only possible in 3D pathology.
%-------------------------------------------------------------------------
\subsection{Inter-slice pooling for 2.5D integration}\label{sec:inter-slice}

% First, $m$ images above and below $\mathbf{X}_k$ within a certain distance, denoted as $\{\mathbf{X}_{k-m}, ..., \mathbf{X}_k,..., \mathbf{X}_{k+m}\}$, are converted into image-level representations through a sequence of patching, feature extraction, and attention-based multiple-instance learning framework.

To leverage contextual information from neighboring images and improve the prediction of the SOI $\mathbf{X}_k$, we incorporate $m$ additional slices above and below $\mathbf{X}_k$ with the spacing of $d$ slices in between, i.e., $\{\mathbf{X}_{k+id}\}_{i=-m}^m$. All of these slices are converted to slice-level features following the steps in Section~\ref{sec:encoder} and~\ref{sec:intra-slice}. 

We subsequently apply weighted averaging on the SOI and neighboring slice-level features, where the weights are learnable, to generate a context-aware SOI feature $\mathbf{\widetilde{z}}_k$ for the downstream prediction task. 
Specifically, we introduce $\mathbf{L}\in\mathbb{R}^{512}$ to learn slice-level weights $\{r_i\}_{i=-m}^{m}$ such that
\begin{equation}
  r_i = \frac{\exp\{\mathbf{L}^T \mathbf{z}_{k+id}\}} {\sum_{i=-m}^{m} \exp\{\mathbf{L}^T \mathbf{z}_{k+id}\}}.
  \label{eq:seqattn}
\end{equation}
Similar to attention-based weights in intra-slice aggregation, the weights $\{r_i\}_{i=-m}^{m}$ emphasize diagnostically-relevant slices. Nevertheless, since this module deals with only a few slice-level features that are already more discernible than patch-level features, we use $\mathbf{L}$, which has far fewer parameters than Eq.~\ref{eq:attnscore}.
The feature $\mathbf{\widetilde{z}}_k$ is then calculated as the weighted average of the slice features,
\begin{equation}
  \mathbf{\widetilde{z}}_k = \sum_{i=-m}^{m} r_i \mathbf{z}_{k+id}.
  \label{eq:seqagg}
\end{equation}
Here $r_i$ refers to the weight of $\mathbf{z}_{k+id}$, which are summed up to 1. $\mathbf{L}$ is built to learn to weigh each slice-level feature, $\mathbf{L} \in \mathbb{R}^{512}$ and its parameters are learned during training.
    
% \textcolor{red}{Rather than aggregating from the union of the set of patch features across SOI and its neighboring slices to form $\mathbf{\widetilde{z}}_k$, we hypothesize that confining the attention network within 2D images enables more effective identification of important patches from a smaller set.}

%-------------------------------------------------------------------------
\subsection{Classification module}\label{sec:clf}

As a final step, the prediction $\widehat{y}_k$ on $\mathbf{X}_k$ is calculated as follows. Here $\mathbf{C}$ is the classification layer, $\mathbf{C} \in \mathbb{R}^{512 \times n}$, and $n$ is the number of risk classes. $n=2$ for this study.

\begin{equation}
  \widehat{y}_k=\operatorname{softmax}(\mathbf{C}^T \mathbf{\widetilde{z}}_k + b).
  \label{eq:classification}
\end{equation}

% The above pipeline can be applied to all image sections in a 3D pathology dataset to produce per-level predictions. Highest risk image sections can be subsequently selected for pathologist evaluation.

\subsection{Training and inference}\label{sec:training}
For the training phase, we select one or two representative slices from each biopsy volume (typically the center slice) as the input and its corresponding pathologists-provided label as the target (\textbf{Figure~\ref{fig:model}(b)}). Since ground truth generation for all slices in each biopsy is infeasible, we extracted representative slices from each 3D dataset to form the training set, promoting the inclusion of images from more diverse patients and biopsies.
% During the training phase, we utilized selected slices with their associated context extracted from 3D data to teach our model to predict corresponding slice-level labels. Note that using all slices from 3D pathology data for training was impractical due to the considerable time required for pathologists to label them. 
% As a result, we extracted a subset of slices from each 3D dataset to form a diverse training set, promoting the inclusion of images from various patients and biopsies.The limited use of slices is influenced by the considerable time required for pathologists to label each section, which makes ground truth generation for all slices in each biopsy impossible.
During the inference phase, the trained model can be deployed across all image sections to generate a predicted risk profile for the volume at the axial sampling pitch (1$\mu m$) of each slice (\textbf{Figure~\ref{fig:model}(c)}). In a clinical context, the image sections with the highest risks would be selected for further evaluation by pathologists.

\section{Experiments}
%-------------------------------------------------------------------------
\subsection{Data description}
We implemented $\ours$ for prognostic risk stratification of prostate cancer based on Gleason gradings provided by a panel of 6 board-certified genitourinary pathologists. Simulated core-needle biopsies (roughly $1 \times 1 \times 15 mm$) were cut from cancer-enriched regions of archived prostatectomy specimens that were previously formalin-fixed and paraffin-embedded (FFPE). The biopsies were stained with a fluorescent analog of hematoxylin and eosin, optically cleared to make them transparent, and imaged with OTLS microscopy in 3D to generate 16-bit gray-scale datasets of two fluorescent channels (nuclear stain and cytoplasmic stain) \cite{protocol,otls2}.  The sampling pitch of the images is $\sim$ 1 $\mu m$/pixel at 10X-equivalent optical resolution laterally (the typical magnification used for Gleason grading) and $\sim$ 4 $\mu m$ optical resolution axially (similar to the thickness of standard slide-based histology). 

A total of 124 slices were selected from 115 OTLS-imaged prostate biopsies across 54 patients. These images were false-colored to mimic the appearance of H\&E histology using a physics-based approach \cite{falsecolor}. For prostate cancer, pathologists quantify the aggressiveness of the cancer using the Gleason grading scheme. Grade group 1 (GG=1) is categorized as low-grade prostate cancer, where patients typically opt for active surveillance (i.e. monitoring without aggressive treatment). Grade Group 2 (GG$\geq$ 2) cancer and above is considered intermediate or high-grade and patients typically receive curative therapy. To enable accurate characterization of higher-grade prostate cancer to provide potentially life-saving treatments, we trained our algorithm to classify slices within 3D prostate datasets as containing low-grade (GG = 1) vs. higher-grade (GG $\geq$ 2) prostate cancer (i.e., a binary classification task).

\begin{figure}[h]
  \centering
   \includegraphics[width=0.85\linewidth]{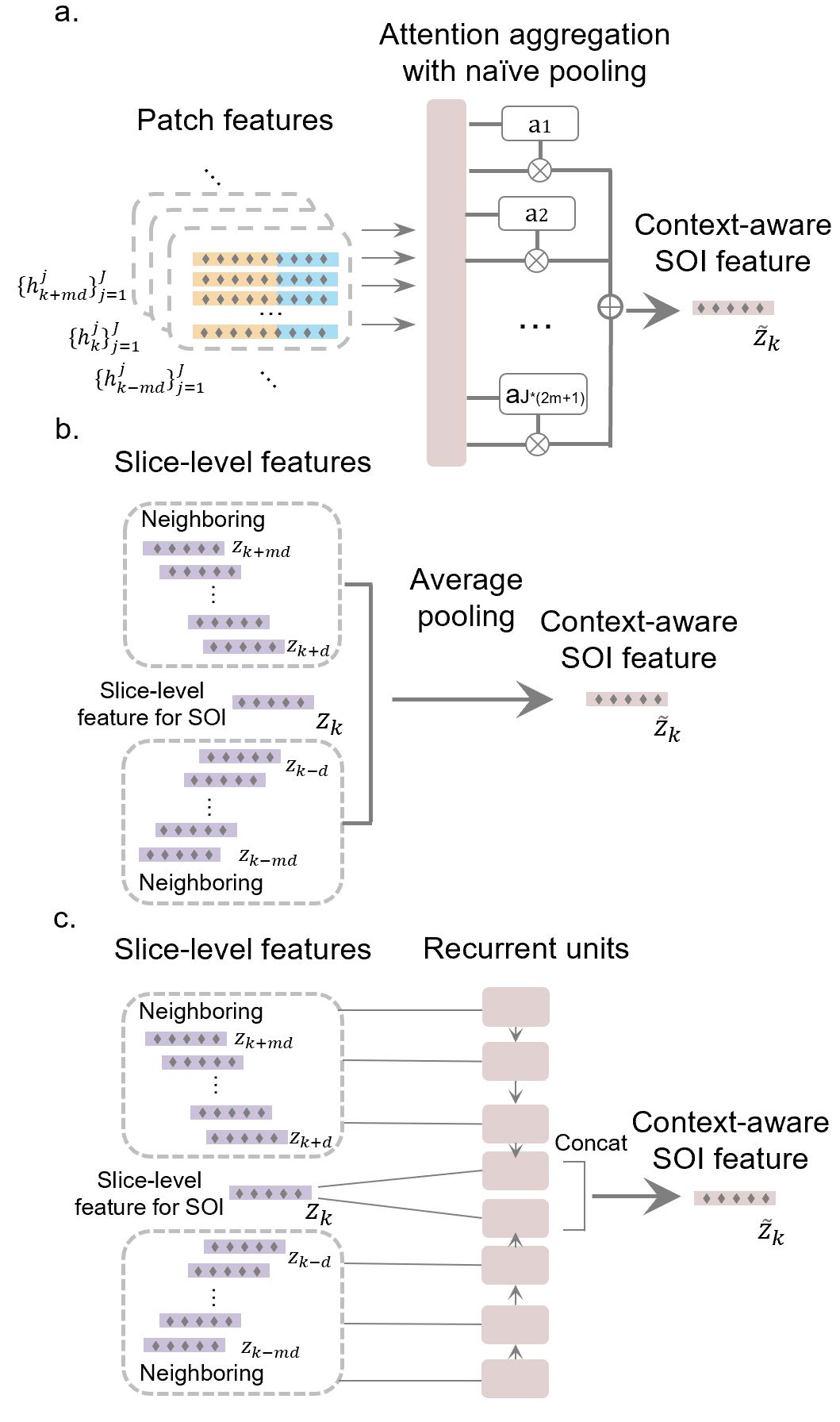}

   \caption{\textbf{Baseline architectures for inter-slice pooling.} a) Naive pooling. b) Average pooling. c) RNN-based pooling.}
   \label{fig:inter}
\end{figure}

\subsection{Implementations}

2D slices were split into non-overlapping 256 $\times$ 256 patches (256 $\times$ 256 $\mu m$).  We used the original two-channel 16-bit gray-scale fluorescence images for training $\ours$. Since the cytoplasm channel exhibits relatively uniform signal distributions, we normalized the 16-bit cytoplasm-channel images after cropping out the background (calculated by Otsu thresholding) and outlier signals (99th percentile). For the nuclear channel, which can vary greatly across an image (depending upon the cellularity of various tissue regions), we performed intensity cropping and normalization on individual patches. Finally, we distributed the normalized nuclear and cytoplasmic channels across the first two channels of the three-channel RGB inputs and left the third channel (B channel) empty.

Dual NVIDIA GeForce RTX 3090 GPUs were used for training and inference. We used the Adam optimizer with the constant learning rate of $2\times 10^{-4}$ and a batch size of 256 to iteratively minimize the cross-entropy loss for all experiments. For 2.5D analysis, neighboring slices within a range of 80 $\mu m$ above and below each image section were selected -- this allows the incorporation of additional information while remaining within the estimated distance of intra-biopsy grading variability~\cite{heter_prostate}. We tested $m\in\{1, 2, 4, 8\}$, maintaining $md=80$, and chose best-performing $m$ for each baseline. 
% For ABMIL+recrrent units, the optimal image number is $m=4$, while all others achieved the best performance at $m=1$. Other details can be found at the GitHub site.

For evaluation, we used the Area Under the ROC Curve (AUC). We also reported the F2 score, which prioritizes recall vs. precision in comparison to the F1 score which weights them equally. This choice reflects our study's goal of screening more aggressive prostate cancer patients. The best F2 score is reported by iterating over different thresholds for mapping predicted probabilities to class labels. We employed a leave-one-out cross-validation strategy, where in each fold, slices from one patient are held out to evaluate model performance, and the remaining images are used for training. Predictions were combined across the patients to calculate the cohort-level performance metrics. 

% \begin{equation}
%   F2 = \frac{5 \times TP} {5 \times TP + FP + 4 \times TN}
%   \label{eq:f2}
% \end{equation}

% Here, TP, FP, and FN, denote the number of true positives, false positives, and false negatives, respectively. We reported the best F2 score by iterating over different thresholds for mapping predicted probabilities to class labels. 

%-------------------------------------------------------------------------
\subsection{Baselines}

Our baselines consist of variations to two modules: 1) Patch feature encoding and 2) Inter-slice pooling (\textbf{Figure~\ref{fig:model}(a)}). All baselines use ABMIL for intra-slice feature aggregation, a lightweight model specifically selected for the data scale.

\noindent \textbf{Patch feature encoding} We first investigated the most widely-used patch feature encoders, including a ResNet50~\cite{resnet} pretrained on ImageNet ($d=1,024$) and a self-supervised CTransPath (a hybrid CNN / Transformer) pretrained on histopathology images ($d=768$)~\cite{cpath}. For ResNet50 features, we applied patch augmentations based on random flipping and brightness/contrast jittering to increase the feature diversity (ResNet50$_{\text{aug}}$). CTransPath features are not augmented as the self-supervised training of the model makes it invariant to augmentations. Motivated by the success of using multiple features for classification~\cite{nguyen2018deep}, we also tested concatenating the ResNet50 and ResNet50$_{\text{aug}}$ features with CTransPath features ($d=1,792$) for more expressive representations. 

\noindent \textbf{Inter-slice pooling} We explored various approaches for inter-slice context aggregation (\textbf{Figure~\ref{fig:inter}}).

\noindent 1) \textbf{No pooling} Only patch features from $\mathbf{X}_k$ are aggregated, and therefore no contextual information is used. 

\noindent 2) \textbf{Naive pooling} All patch features from $\{\mathbf{X}_{k+id}\}_{i=-m}^m$ are aggregated by a single attention module to construct $\mathbf{\widetilde{z}}_k$, disregarding slice identity. 

\noindent 3) \textbf{Average pooling} Patch features are first aggregated to slice-level features within each image. The neighboring slice-level features are averaged with SOI feature to form $\mathbf{\widetilde{z}}_k$, where $\mathbf{\widetilde{z}}_k = \frac{1} {(2m+1)} \sum_{i=-m}^{m} \mathbf{z}_{k+id}$. 

\noindent 4) \textbf{RNN-based pooling} Since average pooling is insensitive to the order of images, we tested RNN-based pooling to incorporate sequential information. Recurrent units aggregate image sequences $\{\mathbf{X}_{k+id}\}_{i=-m}^m$ in a bi-directional manner, integrating slice-level features into the hidden state $\mathbf{hid}$ as contextual information,
\begin{equation}\label{eq:rnn}
\begin{split}
  \mathbf{hid}_{i,1} &= \operatorname{tanh}(\mathbf{W}_{n} \mathbf{z}_{k+id} + \mathbf{W}_{h} \mathbf{hid}_{i+1,1})\\
  \mathbf{hid}_{i,-1} &= \operatorname{tanh}(\mathbf{W}_{n} \mathbf{z}_{k+id} + \mathbf{W}_{h} \mathbf{hid}_{i-1,-1}),\\
\end{split}
\end{equation}
where $\mathbf{W}_{n}, \mathbf{W}_{h} \in \mathbb{R}^{512 \times 512}$ are weight matrices and +1 or -1 denotes the direction of integration. 
% The initial hidden state feature is set to random gaussian noises. 
We concatenate the hidden state features at level $k$ to form $\mathbf{\widetilde{z}}_k$, where $\mathbf{\widetilde{z}}_k = \mathbf{hid}_{k,1} \oplus \mathbf{hid}_{k,-1}$.
%The two sets of parameters $\mathbf{W}_in \in \mathbb{R}^{512 \times 512}$ and $\mathbf{W}_hid \in \mathbb{R}^{512 \times 512}$ in RNN units dictate what is incorporated from the new input image feature vector and what is kept from the previous hidden state feature vector to incorporate into a new hidden state feature vector. $sign(i-k)$ indicate the direction of integration (-1 or +1).
%-------------------------------------------------------------------------The two predominant Gleason patterns are reported and determine the ISUP grade group (GG). Grade group 1 (Gleason pattern 3+3) is categorized as low-grade prostate cancer, where patients typically opt for active surveillance. The presence of Gleason pattern 4 or 5 is associated with intermediate or high-grade prostate cancer (GG 2 cancer and above), where patients typically receive curative therapy. Since it is paramount to accurately identify and characterize higher-grade prostate cancer (GG $\geq$ 2) to provide potentially life-saving treatments, we trained our algorithm to classify image sections within 3D prostate datasets as containing low-grade (GG = 1) vs. higher-grade (GG $\geq$ 2) prostate cancer (i.e., a binary classification task).

\section{Results}
% We first report the results from ablating patch feature encoders for 2D ABMIL in section 5.1. Based on the observed best patch encoding strategy, we further compare our 2.5D approach with the 2D baseline in section 5.2. Other methods of incorporating contextual information are investigated as well to gain a better understanding of the 2.5D models. Finally, we implement our triage model on an example 3D pathology dataset to illustrate the potential to improve prostate cancer grading compared with standard 2D histology while reducing pathologist workloads (compared with the current standard of care).

%-------------------------------------------------------------------------
\subsection{Patch feature encoders}

\begin{table}
  \centering
  \begin{tabular}{@{}l|cc@{}}
    \toprule
    Encoders & AUC ($\uparrow$) & F2 ($\uparrow$)\\
    \midrule
    \rule{-2pt}{0pt}
    ResNet50 & \shortstack{75.7\% \\(68.2\%-82.6\%)} & \shortstack{87.5\% \\(84.3\%-90.9\%)}\\
    \hline
    \rule{-2pt}{20pt}
    \shortstack[l]{ResNet50$_{\text{aug}}$} & \shortstack{75.6\% \\(67.9\%-82.3\%)} & \shortstack{87.9\% \\(84.5\%-90.9\%)}\\
    \hline
    \rule{-2pt}{20pt} 
    CTransPath & \shortstack{80.1\% \\(73.3\%-86.4\%)} & \shortstack{86.7\% \\(83.6\%-90.3\%)}\\
    \hline
    \rule{-2pt}{20pt} 
    \shortstack[l]{ResNet50 $\oplus$ \\CTransPath} & \shortstack{79.7\% \\(73\%-86.2\%)} & \shortstack{87.0\% \\(83.8\%-90.5\%)}\\
    \hline
    \rule{-2pt}{20pt} 
    \shortstack[l]{ResNet50$_{\text{aug}}$ $\oplus$ \\CTransPath} & \textbf{\shortstack{81.3\% \\(75.2\%-87.3\%)}} & \textbf{\shortstack{88.5\% \\(85.5\%-91.7\%)}}\\
    \bottomrule
  \end{tabular}
  \caption{\textbf{Results for different encoders}. Best performances are in \textbf{bold}. The 95\% confidence interval (95\% CI) is calculated based on bootstrapping. Concatenation is denoted by $\oplus$.}
  \label{tab:patch}
\end{table}

% \begin{table*}[!ht]
%     \centering
%     \begin{tabular}{@{}l|c|cc@{}}
%         \toprule
%         & Aggregation & AUC ($\uparrow$) & F2  ($\uparrow$) \\
%         \midrule
%         \rule{-2pt}{0pt}
%         \shortstack{2D}& None & \shortstack{81.3\% \\(75.2\%-87.3\%)} & \shortstack{88.5\% \\(85.5\%-91.7\%)} \\
%         \rule{-2pt}{20pt}
%         & \shortstack{naive} & \shortstack{87.2\% \\(81.5\%-82.2\%)} & \shortstack{88.2\% \\(85.7\%-92.4\%)} \\
%         \cline{2-4}
%         \rule{-2pt}{21pt}
%         \shortstack{2.5D}& \shortstack{average\\pooling} & \shortstack{89.8\% \\(85.1\%-94\%)} & \shortstack{90.1\% \\(87.5\%-93.6\%)} \\
%         \cline{2-4}
%         \rule{-2pt}{20pt}
%         & \shortstack{recurrent\\units} & \shortstack{89.3\% \\(84.2\%-93.6\%)} & \shortstack{90.5\% \\(87\%-93.6\%)} \\
%         \cline{2-4}
%         \rule{-2pt}{20pt}
%         & \shortstack{weighted\\averaging} & \textbf{\shortstack{90.1\% \\(85.3\%-94.3\%)}} & \textbf{\shortstack{92.4\% \\(90.2\%-95.1\%)}} \\
%         \bottomrule
%     \end{tabular}
%     \caption{\textbf{Results for inter-slice aggregation approaches}. All methods use the concatenation of augmented ResNet50 and CTransPath features. Best performances are indicated in \textbf{bold}. The 95\% confidence interval (95\% CI) is calculated based on bootstrapping.}
%     \label{tab:25D}
% \end{table*}

The results for feature encoder ablation can be found in Table~\ref{tab:patch}. We observe that the concatenation of augmented ResNet50 and CTransPath features outperformed all other strategies using 2D attention-based MIL.  
We observe that ResNet50 features and augmented ResNet50 features achieved a similar AUC, but augmentation helped slightly with identifying the positive class (better F2 score). CTransPath features improved AUC by a significant margin. However, the F2 score is inferior to experiments using original and augmented ResNet50 features, which indicate the improved AUC mainly comes from correctly predicting GG=1 images, while many GG$\geq$2 images are still confused with the low-grade class (i.e. poor recall or diagnostic sensitivity).  Concatenation of ResNet50 and CTransPath features results in a better F2 score than CTransPath, but at the expense of slightly lower AUC. Applying patch augmentation on ResNet50 encoders and concatenating the features with CTransPath features achieved the best AUC and F2 score compared with all other experiments. Therefore, we used these features for subsequent analyses. We leave utilizing the patch features from very recent foundation models as future work~\cite{lu2024visual, chen2024towards}.

%-------------------------------------------------------------------------
\begin{table}[!ht]
    \centering
    \begin{tabular}{@{}c|cc@{}}
        \toprule
        Aggregation & AUC ($\uparrow$) & F2  ($\uparrow$) \\
        \midrule
        \rule{-2pt}{0pt}
        None & \shortstack{81.3\% \\(75.2\%-87.3\%)} & \shortstack{88.5\% \\(85.5\%-91.7\%)} \\
        \hline
        \rule{-2pt}{20pt}
        \shortstack{Naive} & \shortstack{87.2\% \\(81.5\%-82.2\%)} & \shortstack{88.2\% \\(85.7\%-92.4\%)} \\
        \hline
        \rule{-2pt}{21pt}
        \shortstack{Average} & \shortstack{89.8\% \\(85.1\%-94\%)} & \shortstack{90.1\% \\(87.5\%-93.6\%)} \\
        \hline
        \rule{-2pt}{20pt}
        \shortstack{RNN} & \shortstack{89.3\% \\(84.2\%-93.6\%)} & \shortstack{90.5\% \\(87\%-93.6\%)} \\
        \hline
        \rule{-2pt}{20pt}
        \shortstack{Weighted\\Average} & \textbf{\shortstack{90.1\% \\(85.3\%-94.3\%)}} & \textbf{\shortstack{92.4\% \\(90.2\%-95.1\%)}} \\
        \bottomrule
    \end{tabular}
    \caption{\textbf{Results for inter-slice aggregation approaches}. All methods use the concatenation of ResNet50$_{\text{aug}}$ and CTransPath features. Best performances are in \textbf{bold}. The 95\% confidence interval (95\% CI) is calculated based on bootstrapping.}
    \label{tab:25D}
\end{table}

\subsection{Performance comparison between 2.5D vs 2D}
The comparisons between 2D and 2.5D approaches can be found in Table~\ref{tab:25D}.
We observe that 2.5D methods using contextual information improved the classification of low-grade vs. higher-grade cancer over 2D analysis. We considered attention-based MIL on independent 2D images with no inter-slice aggregation as the baseline and explored different inter-slice context aggregation strategies. The \textit{naive pooling} baseline, where a single attention-based network aggregates patches from the SOI and neighboring slices, results in notable improvement in AUC, suggesting that additional spatial context provided by the 3rd dimension enhances the discriminative capabilities of the model. However, there was a drop in F2 (poorer recall or diagnostic sensitivity), potentially due to the increased number of patches to attend to, making it more challenging to correctly recognize positive patches containing small foci of aggressive cancer.

To better utilize the depth information within each biopsy, we further analyzed strategies for a sequence of aggregations, first within each slice (ABMIL) followed by across slices.  For inter-slice aggregation, we observe that simple averaging of slice features (\textit{average pooling}) achieves further improvements in both AUC and F2. Interestingly, the \textit{RNN-based pooling}, designed to better incorporate sequential information from neighboring images, performs on par with the average pooling. We conjecture that the introduction of more trainable parameters with RNN likely leads to overfitting. Finally, since one limitation of average pooling is that positive images may still be diluted by neighboring images, especially when aggressive cancer constitutes only a small proportion, we designed \textit{weighted averaging} to emphasize slice features that are more diagnostically important. We observe that weighted averaging achieved a significant improvement in AUC and especially in F2. We conclude that a sequence of aggregation based on weighted averaging employed by $\ours$ can best utilize the 3D dataset for clinical prediction.

%-------------------------------------------------------------------------
\subsection{Interpretability}
\begin{figure*}[!h]
  \centering
   \includegraphics[width=0.8\linewidth]{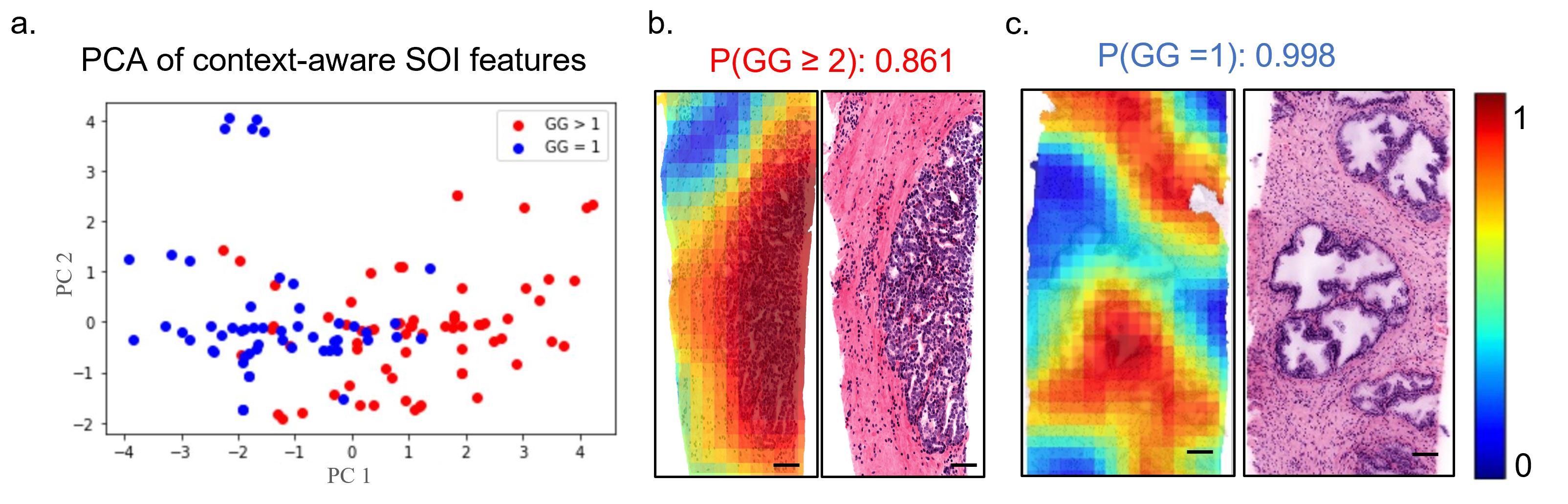}

   \caption{\textbf{Visualization of SOI features and interpretable attention heatmaps.} a) PCA of context-aware SOI features. b) and c) show examples of attention heatmaps with corresponding false-colored images. The scale bar is 100 $\mu m$. Color bar indicates attention scores.}
   \label{fig:interp}
\end{figure*}

To gain a better understanding of the $\ours$ predictions, we applied principal component analysis (PCA) to the context-aware SOI features in \textbf{Figure~\ref{fig:interp}(a)}. GG $\geq$ 2 slices are visually separable from GG = 1 slices in the PC space, suggesting that our model can discriminate between two classes. By visualizing heatmaps of attention scores overlaid on top of false-colored H\&E-like images, we can identify regions important for rendering predictions. For example, regions in \textbf{Figure~\ref{fig:interp}(b)} are predicted as highly likely to contain higher-grade prostate cancer. The high-attention regions are roughly aligned with fused glands, which are commonly associated with higher-grade prostate cancer. Regions in \textbf{Figure~\ref{fig:interp}(c)} are predicted as low-grade prostate cancer, where the highly attended regions correspond to well-formed benign glands and lymphocytes.

\begin{figure*}[!h]
  \centering
   \includegraphics[width=1.0\linewidth]{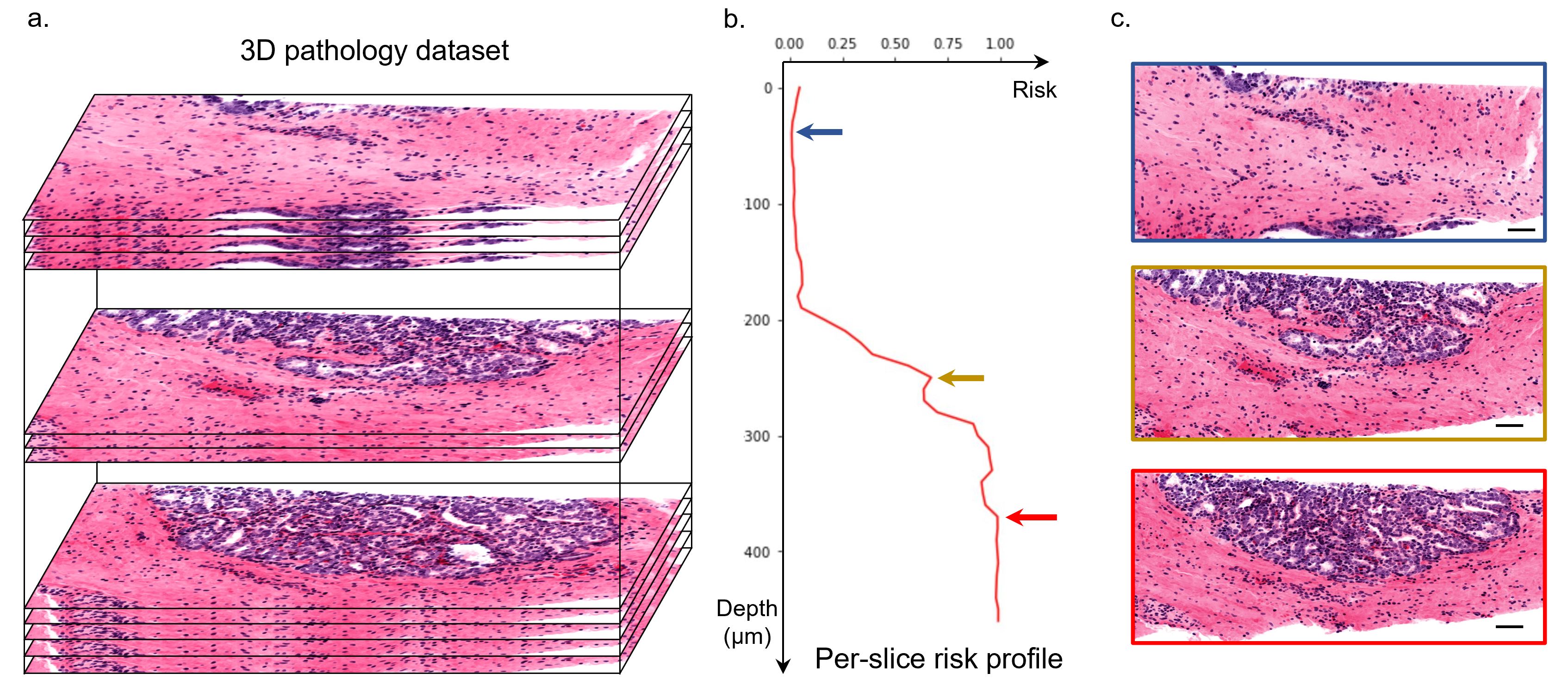}

   \caption{\textbf{$\ours$ triage on an example 3D pathology dataset.} a) An example 3D pathology dataset. b) Per-slice risk profile, predicted by $\ours$, for higher-grade prostate cancer. c) Images at arrow positions are reviewed by a board-certified pathologist, showing that human evaluation on select slices broadly aligns with the risk profile. The scale bar is 100 $\mu m$.}
   \label{fig:profile}
\end{figure*}
\subsection{Triage on a 3D pathology dataset}
We demonstrate the slice-by-slice inference of the trained $\ours$ on an example 3D dataset (\textbf{Figure~\ref{fig:profile}(a)}). Specifically, we generate a depth profile of predicted context-aware risk scores, with the risk defined as the predicted probability of a given slice containing higher-grade prostate cancer. To validate our findings, a board-certified pathologist reviewed images at the depths indicated by arrows (\textbf{Figure~\ref{fig:profile}(b)}). At the blue arrow, only well-formed benign glands are seen (Grade Group 1), albeit with only a small number of glands visible at this depth. Fused glands (associated with higher-grade prostate cancer) are present in the slice at the brown arrow, but benign glands are more prevalent (Grade Group 2). The fused glands become the predominant morphology at the red arrow, suggesting more aggressive cancer (Grade Group 3). In summary, the pathologist evaluation of select slices broadly aligns with the risk profile generated by our model. In real-world clinical practice, the highest risk slice at the red arrow as in \textbf{Figure~\ref{fig:profile}(c)} would be prioritized for pathologist review. A large-scale clinical study will be performed in the future to demonstrate the ability of AI-triaged 3D pathology (enabled by $\ours$) to improve the detection of higher-grade prostate cancer in comparison to standard 2D histopathology.

%Here, we refer to images containing intermediate-to-high grade prostate cancer (GG$\geq$2) as the positive class (positive for more aggressive cancer) while images with only low-grade prostate cancer (GG=1) are considered as the negative class (negative for more aggressive cancer).
\section{Conclusion}
We present $\ours$, a 2.5D multiple instance learning framework to triage the highest-risk slices within 3D pathology datasets to facilitate pathologist review. Our work leverages contextual information in 3D pathology data to enhance the predicting accuracy of each slice. $\ours$ could potentially accelerate the clinical adoption of 3D pathology by improving pathologists' diagnostic accuracy via increased tissue sampling and context-aware triage. Future work includes curation of more 3D pathology data across organs for large-scale clinical validation in comparison to standard 2D histopathology. 
%compared to the current clinical practice of examining only a few 2D tissue sections. With \ours utilizing contextual information brought forth by the addition of the third dimension, we believe we have come closer to the clinical adoption of 3D pathology.In particular, our final goal is to show that AI-triaged 3D pathology can significantly improve diagnostic determinations compared to standard-of-care 2D histopathology while also reducing pathologist workloads, a double advantage that should drive rapid clinical adoption of this low-risk strategy that keeps pathologists “in the loop.”

\clearpage
{
    \small
    \bibliographystyle{ieeenat_fullname}
    \bibliography{main}
}


\end{document}